# How Covid-19 Pandemic Changes the Theory of Economics?

Matti Estola

University of Eastern Finland, Faculty of Social Sciences,
P.O. Box 111, FIN-80101 Joensuu, Finland.
E-mail: matti.estola@uef.fi

**Abstract.** *During its history, the ultimate goal of economics has been to develop similar frameworks for modeling economic behavior as invented in physics. This has not been successful, however, and current state of the process is the neoclassical framework that bases on static optimization. By using a static framework, however, we cannot model and forecast the time paths of economic quantities because for a growing firm or a firm going into bankruptcy, a positive profit maximizing flow of production does not exist. Due to these problems, we present a dynamic theory for the production of a profit-seeking firm where the firm's adjustment may be stable or unstable. This is important, currently, because we should be able to forecast the possible future bankruptcies of firms due to the Covid-19 pandemic. By using the model, we can solve the time moment of bankruptcy of a firm as a function of several parameters. The proposed model is mathematically identical with Newton's model of a particle moving in a resisting medium, and so the model explains the reasons that stop the motion too. The frameworks for modeling dynamic events in physics are thus applicable in economics, and we give reasons why physics is more important for the development of economics than pure mathematics. (JEL D21, O12)*

## 1. Introduction

Mirowski [1] shows that aim of the progenitors of neoclassical economics was to create similar principles of modeling for economics as in classical mechanics. This process has not been successful, however, and current theoretical framework of neoclassical economics relies on static optimization and the concept of equilibrium borrowed from physics by Canard at 1801, see [2].

By a static model, however, we cannot explain the evolutionary behavior of economies. MasColell et al. [3 p. 620] state this problem as: *"A characteristic feature that distinguishes economics from other scientific fields is that, for us, the equations of equilibrium constitute the center of our discipline. Other sciences, such as physics or even ecology, put comparatively more emphasis on determination of dynamic laws of change. ... The reason, informally speaking, is that economists are good (or so we hope) at recognizing a state of equilibrium but are poor at predicting precisely how an economy in disequilibrium will evolve. Certainly there are intuitive dynamic principles: if demand is larger than supply then price will increase, if price is larger than marginal cost then production will expand. ... The difficulty is in transforming these informal principles into precise dynamic laws."*

The definition of *"dynamic economic laws"* requires a dynamic framework for modeling, and the Newtonian framework presented in Refs [4-8] is suitable for this. Economic units like to improve their welfare and optimal behavior results when no improvement is possible, see [9]. *If* we thus assume that *economic units (entrepreneurs, consumers, workers, etc.) like to better their situation in* *a competitive environment,* the units behave as particles in the physical world that "like" to move toward the position of their minimum potential energy. *The error in the neoclassical framework is the same as if in physics the initial position of a particle were its point of minimum potential energy where it has "no willingness" to move anywhere.*

Due to the Covid-19 pandemic, thousands of firms around the world will go into bankruptcy in the near future. Statistics Finland (https://www.stat.fi/index_en.html) e.g. reports that in Finland during Jan-Oct 2020 were initiated 1886 processes of bankruptcy. We show that we can forecast this phenomenon by using the Newtonian framework of economics in contrast with the neoclassical one.

## 2. Weaknesses in the neoclassical theory of a firm

Let the flow of production of a firm under imperfect competition be $q$ $(unit/y)$, where $y$ is a time unit e.g. a week, month, year, etc[1]. The sales function of the one-product firm is denoted by $p = f(q)$, where $p(€/unit)$ is the price of the product. The cost function of the firm is $C(q)$ $(€/y)$ and the profit of the firm $\Pi(q)$ $(€/y)$ from time unit $y$ is then

$$\Pi(q) = f(q)q - C(q),\ f'(q) < 0,\ C'(q) > 0.$$

The necessary condition for maximum profit assumed in the neoclassical theory is

$$\Pi'(q) = 0 \iff f(q) + f'(q)q = C'(q), \qquad (1)$$

where $MR = f(q) + f'(q)q$ is marginal revenue and $MC = C'(q)$ marginal cost. The possible solution of Eq. (1) is denoted by $q^*$. The sufficient condition for $q^*$ to maximize the profit is

$$\Pi''(q)|_{q=q^*} = 2f'(q) + f''(q)q - C''(q) < 0. \qquad (2)$$

Condition (2) requires that for the sales function holds

$$2f'(q) + f''(q)q \leq 0;$$

we will return to this condition in section 5.1. The requirement of maximum profit for the cost function is $C''(q) > 0$, that is, marginal costs are increasing with $q$, i.e. decreasing returns to scale hold. Increasing returns to scale takes place in a firm e.g. when small-scale production is changed to larger scale production via automatizing and producing in larger quantities. We have evidence of this (see [10-13]) and thus the requirement $C''(q) > 0$ implies that the static neoclassical framework does not cover these situations. In the real world, the sales and cost functions of firms depend on time, which too is impossible in the static neoclassical framework.

Due to these weaknesses in the neoclassical framework, we introduce a general framework for modeling the behavior of a firm that can explain besides static optimal behavior also its growth and possible bankruptcy. In the theoretical part, we follow [6] and so the presentation is relatively short.

## 3. Why physics is more important for economics than pure mathematics?

[1] Measurement units are in parentheses after the quantities.

Textbooks of economics commonly refer to those of mathematics when presenting mathematical theories. However, there are four reasons why it would be more fruitful to refer to books of physics.

1) Mathematics operates with numbers, vectors, functions, and other abstract objects because in mathematics, measurement units are not defined. Real sciences like physics and economics, on the other hand, operate with dimensional quantities (i.e. quantities with a measurement unit) measured from the real world. Thus, for empirical evaluation it is essential that we express the mathematical theories in economics with dimensional quantities as in physics because then the theoretical quantities have exact counterparts in the real world and the theories are rigorously testable. For example, in dynamic models mathematics operates with derivatives while time derivatives of theoretical physical and economic quantities correspond to instantaneous flows or accelerations of these quantities with exact measurement units.

2) Mathematics studies the principles of solving equations and differential equations, the algebraic structures of mathematical objects, and the properties of functions and functional spaces. The modeling of real world events, on the other hand, requires the measurement of flows and accelerations of physical, economic, etc. quantities, and mathematical formulation of reasons causing these changes. In mathematics, events are not observed, and the causal modeling principle in real sciences – every event has a reason – is not applied in mathematics. Physics has shown how to apply the causal principle – force causes acceleration – successfully in explaining real world events, and other real sciences can apply the same principle.

3) In mathematics, a theorem has scientific value after it has been proved formally. In real sciences, too, a theory should not have inconsistencies. However, in real sciences the scientific value of a theory depends on its ability to explain observed events. Thus, in real sciences a consistent theory does not have scientific value unless it has been corroborated empirically.

4) One of the “laws of Nature” is that freely moving particles move along the shortest path toward the direction where their potential energy decreases[2]. Lagrangian mechanics in physics bases on this principle, and there the equilibrium state of a particle is its point of minimum potential energy. However, Newtonian mechanics yields the same equations of motion for moving particles as Lagrangian mechanics, and Newtonian mechanics bases on the forces that “guide” particles toward the zero-force state where the minimum potential energy holds. In neoclassical microeconomics, the principle of modeling is that economic units (consumers, firms, etc.) make decisions to reach their optimum state, i.e. maximum utility or profit. Thus, if we define “economic forces” so that they “guide” economic units toward their optimum state, the two sciences are homomorphic and can apply the same principles in modeling the time paths of “moving objects”. The potential “energy” (in economics, money corresponds to energy, see e.g. [18]) in a firm’s production can be defined e.g. as the profitability difference in the future by producing at current flow of production as compared with the profit maximizing one. Thus, a firm has potential “energy” if it can change its flow of production in the future so that its profit increases. In [18], this way defined production potential in the Lagrangian framework gives identical results as the Newtonian framework we apply here. Thus, the two frameworks are consistent in economics as in physics. In mathematics, no general modeling principles exist for observed events, and mathematics helps in these analyses only in checking that no errors occur in calculations.

[2] Think about water moving down along the surface of a hill.

## 4. Kinematics of production

The accumulated production of a firm till time moment $t$ (like the accumulated kilometers a car has driven) denoted by $Q(t)$ $(unit)$ (a marginal change in time $ds$ is measured in time units $y$) is

$$Q(t) = Q(t_0) + \int_{t_0}^{t} q(s)ds, \quad Q'(t) = q(t), \quad Q''(t) = q'(t),$$

where $Q(t_0)$ is the accumulated production of the firm from its foundation till moment $t_0$, $Q'(t)$ with unit $unit/y$ is the instantaneous velocity (flow) of accumulated production, and $Q''(t) = q'(t)$ with unit $unit/y^2$ the instantaneous acceleration of accumulated production. This kinematics of production is a necessary prelude for production dynamics analogous with Newtonian mechanics. Notice that in mathematics, the above concepts are abstract derivatives of functions without a measurement unit and no counterpart in the real world.

## 5. A dynamic theory of production

In dynamizing the theory of a firm, we let the price and the flow of production to depend on time $t$ with unit $y$. The function, that expresses the maximum unit price $p(t)$ $(€/unit)$ by which the firm can sell $q(t)$ $(unit/y)$ units at time unit $y$, the inverse demand or the sales function of the firm, is

$$p(t) = f(q(t), t), \quad \frac{\partial f}{\partial q} \leq 0.$$

We assume that the product is heterogeneous with those of other firms to avoid taking account other firms' production decisions in the decision of the studied firm, see [15]. Including time in function $f(.)$ allows that the sales of the firm may change with time due to changes in customers' habits or due to the firm's marketing of the good.

We can write the cost function of the studied firm as

$$C(q(t), t) = h(t) + g(q(t), t)q(t)$$

where $C(.)$ $(€/y)$ is the costs of the firm during time unit $y$ at time moment $t$ realized at the flow of production $q(t)$, $h(t)$ $(€/y)$ fixed costs during time unit $y$ independent of the flow of production, and unit costs $g(.)$ $(€/unit)$ may depend on the flow of production and time. Technological development e.g. may decrease unit costs with time.

The firm's leaders like to increase the firm's profit with time and so they change $q(t)$ as (see [6]):

$$q'(t) > 0 \;\; if \;\; \frac{\partial \Pi}{\partial q} > 0, \quad q'(t) < 0 \;\; if \;\; \frac{\partial \Pi}{\partial q} < 0, \quad and \quad q'(t) = 0 \;\; if \;\; \frac{\partial \Pi}{\partial q} = 0. \qquad (3)$$

These rules guide the firm toward the direction of greater profit with time. They are in accordance with the intuitive dynamic rules stated in [3] referred earlier, and [16 p. 387] writes about a firm in a perfectly competed market: *"… if $p - \Delta C/\Delta q > 0$, then the firm can increase its profits by producing more"*. Thus, also in neoclassical economics the adjustment rules in Eq. (3) are accepted.

We identify $\partial\Pi/\partial q = MR - MC$, the reason for the acceleration of production, as the *"force acting upon the production of a profit-seeking firm"*. A relation, that fulfills the rules in Eq. (3) (see [6]), is

$$mq'(t) = \frac{\partial\Pi}{\partial q}, \qquad m \geq 0, \qquad \text{constant.} \tag{4}$$

Because constant $m$ is positive or zero, then in Eq. (4), $q'(t)$ and $\partial\Pi/\partial q$ are simultaneously positive, negative, or zero, and so the requirements in Eq. (3) are met. In Eq. (4), the firm adjusts its flow of production according to the deviation between its marginal revenue and cost (this holds in [17] too). The magnitude of $m$ measures the inertia in the firm's adjustment of production, and its unit $(€ \times y^2)/unit^2$ guarantees the dimensional consistency of Eq. (4), see [18]. We call $m$ the *"inertial mass of the flow of production"* and Eq. (4) the *Newtonian theory of a firm adjusting its flow of production*. Neoclassical theory corresponds to zero-force in Eq. (4): $\partial\Pi/\partial q = 0 \Leftrightarrow q'(t) = 0$.

### 5.1. Decreasing and increasing returns to scale

To start up with a simple model, we assume the following sales and cost functions for the firm:

$$p(t) = f\big(q(t)\big) = a + \tfrac{b}{q(t)}, \qquad C(q(t), t) = h(t) + \left(A + \tfrac{B}{2}q(t)\right)q(t), \tag{5}$$

where $q(t)$ is as before and $a, A$ are positive constants with unit $€/unit$, constant $b$ has unit $€/y$, and that of $B$ is $(€ \times y)/unit^2$. If $q(t) \to 0$ then $p(t) \to \infty$, and if $q(t) \to \infty$ then $p(t) \to a$. Parameter $a$ measures the price the firm receives from one product in any situation, and $b$ measures the relation between price and the sales of the firm. The more competition exists, the closer to zero is $b$; monopolistic competition and a monopoly firm correspond to $b > 0$. In section 2 we showed that the sufficient condition for maximum profit of the studied firm is

$$\Pi''(q)|_{q=q^*} = 2f'(q) + f''(q)q - C''(q) < 0.$$

Applying the sales function in Eq. (5) in this, we get

$$2f'(q) + f''(q)q = -2\frac{b}{q^2} + \frac{2b}{q^3}q = 0.$$

Thus, the sufficient condition for maximum profit in this case is $C''(q) > 0$. In the cost function, parameter $A > 0$ measures unit costs from the first produced unit during time unit $y$ and parameter $B$ the relation unit costs have with the flow of production. Decreasing returns to scale correspond to $B > 0$, constant returns to scale to $B = 0$, and increasing returns to scale to $B < 0$.

Assuming fixed costs during time unit $y$ to be constant $h_0$, profit function $\Pi(€/y)$ becomes

$$\Pi\big(q(t)\big) = p\big(q(t)\big)q(t) - C(q(t), t) = aq(t) + b - h_0 - Aq(t) - \tfrac{B}{2}q^2(t). \tag{6}$$

Omitting the time dependency in variable $q$ we get the static neoclassical setting. The necessary condition for maximum profit is then

$$\Pi'(q) = 0 \quad \Leftrightarrow \quad a - A - Bq = 0 \quad \Leftrightarrow \quad q^* = \tfrac{a-A}{B},$$

where $q^*$ is the possible profit maximizing flow of production. Notice that the unit of $q^*$ is: $\frac{€/unit}{€\times y/unit^2} = \frac{unit}{y}$. The sufficient condition for maximum is

$$\Pi''(q)|_{q=q^*} = -B < 0. \quad (7)$$

Condition (7) requires $B > 0$, and for $q^*$ to be positive, then $a > A$ must hold too. If these conditions hold, $q^*$ is the profit maximizing flow of production.

In the following in this section, we assume $a > A$ because producing a good that does not get sold at a price higher than its unit costs in the beginning of production is in most cases not sensible. The marginal revenue and cost functions with unit $€/unit$ are then:

$$MR = a, \quad MC = A + Bq.$$

Increasing returns to scale ($B < 0$) with constant $MR$ then implies that the firm does not have a positive optimal flow of production, and these situations cannot be modeled by using the static neoclassical framework. Next, we show that we can model this situation in the proposed framework.

Newtonian Eq. (4) is now

$$\frac{\partial \Pi}{\partial q} = mq'(t) \iff a - A - Bq(t) = mq'(t). \quad (8)$$

Eq. (8) shows that the firm increases its flow of production ($q'(t) > 0$) if $q(t) < (a - A)/B = q^*$, and decreases its flow of production if $q(t) > q^*$, where $q^*$ is the possible profit maximizing flow of production. The solution of Eq. (8) is

$$q(t) = \frac{a - A}{B} + H_0 e^{-\frac{B}{m}t}, \quad (9)$$

where $H_0$ ($unit/y$) is the constant of integration and $Bt/m$ a dimensionless quantity because time $t$ has unit $y$. Let $B > 0$. Then, according to Eq. (9), $q(t) \to q^* = (a - A)/B$ with $t \to \infty$ and this neoclassical equilibrium corresponds to zero force in Eq. (8). Setting $t = 0$ in Eq. (9) gives: $q(0) = (a - A)/B + H_0 \Longrightarrow H_0 = q(0) - (a - A)/B$. Thus, $q(t)$ increases (decreases) with time if $H_0$ is negative (positive), that is, if $q(0)$ is smaller (greater) than $q^*$. The neoclassical theory corresponds to $H_0 = 0 \Leftrightarrow q(0) = (a - A)/B = q^*$ and then the firm produces at constant flow $q^*$ through time.

This example adds dynamics in the neoclassical framework and shows how the firm reaches its optimum with time if it does not be in it, see Fig. 1a where $q^* = 1000$. Notice the faster convergence in the time path that starts closer to equilibrium. In Fig. 1b is displayed three paths from the “old equilibrium” $q^* = 1000$ to the “new equilibrium” $q^* = (150 - 20)/0.08 = 1625$ that results when $a$ changes from $100 \to 150$. Notice the faster convergence with smaller “inertial mass”. Thus, the model explains how the firm enters from one equilibrium to another, and the unrealistic instantaneous adjustment – assumed in the static neoclassical theory – is obtained with $m \to 0$.

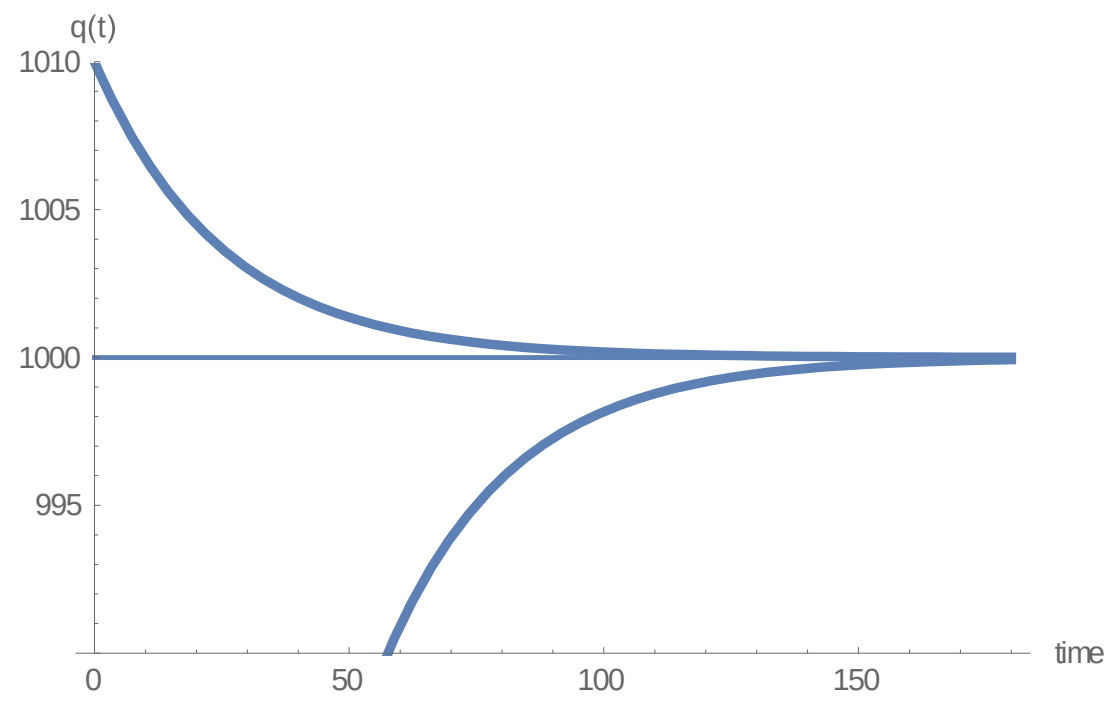

Figure 1a. Eq. (9) with $m = 2,\ a = 100,\ A = 20,\ B = 0.08,\ H_0 = -100,\ and\ H_0 = 10.$

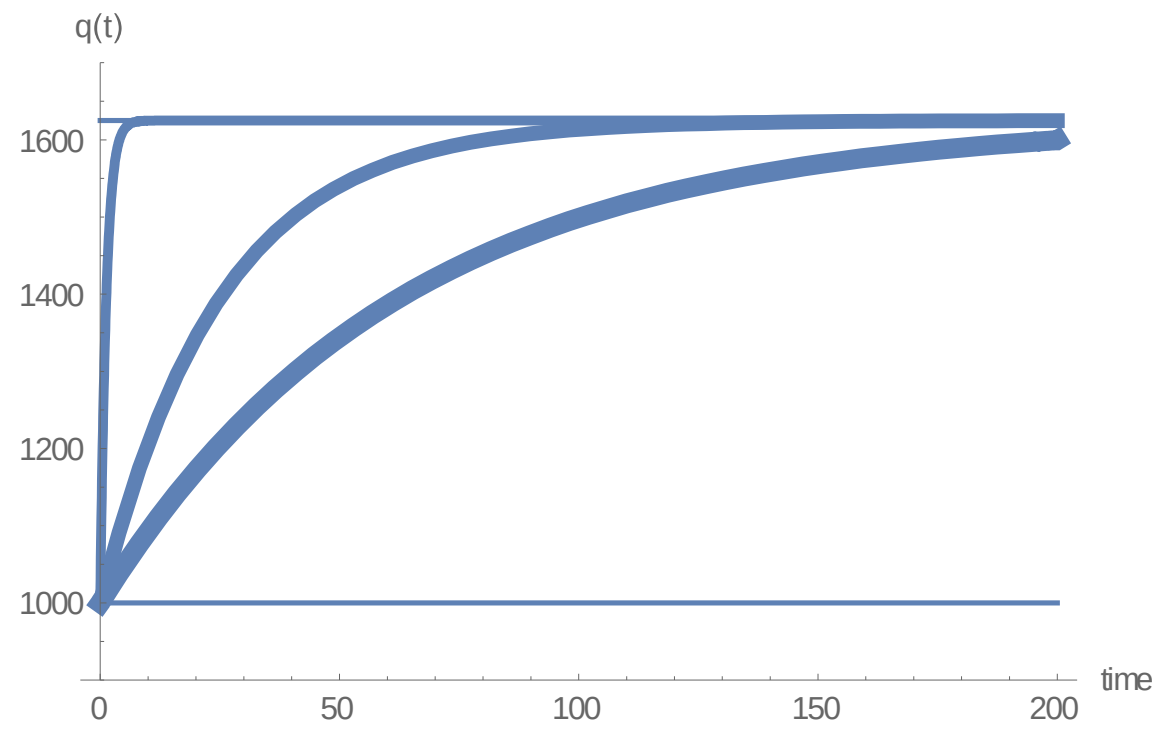

Figure 1b. Eq. (9) with $a = 150,\ A = 20,\ B = 0.08,\ H_0 = 1000 - 1625 = -625,\ m = 0.1,\ m = 2,\ and\ m = 5.$

Suppose next $a - A > 0$ and $B < 0$. Then $q^* < 0$. Because $q(0) \geq 0$ then $H_0 > 0$, and a positive exponential time trend exists in $q(t)$. Thus, $a - A > 0$ with increasing returns to scale leads to an exponential growth in $q(t)$ of the profit-seeking firm. Then a positive equilibrium flow of production does not exist and $q^*$ corresponds to the minimum profit, see Fig. 2a where $q^* = -20$. The equilibrium in Eq. (9) is obtained letting $t \to -\infty$, even though $q(t) < 0$ is not meaningful in economics. The problem here is that increasing $q$ makes unit costs negative, which is impossible.

To correct the problems in the unstable case, suppose the unit cost function is made of two parts:

$$g\big(q(t)\big) = \begin{cases} 90 - (0.5/2)\, q(t), & 0 < q(t) \leq 200, \\ 20 + (0.08/2) q(t), & q(t) > 200. \end{cases}$$

The two parts represent two phases of the production process. The decreasing unit cost part takes place in the beginning of the process, and unit costs start to increase when $q > 200\ (unit/y)$. The flow of production according to Eq. (9) is displayed in Figs. 2a,b, the cost function according to Eq. (5) in Figs. 3a,b, and the profit according to Eq. (6) in Figs. 4a,b; $h_0 = 2000$, and in the first case $H_0 = 20$ and in the latter case $H_0 = -2$. The optimal flow of production is $q^* = 1000$.

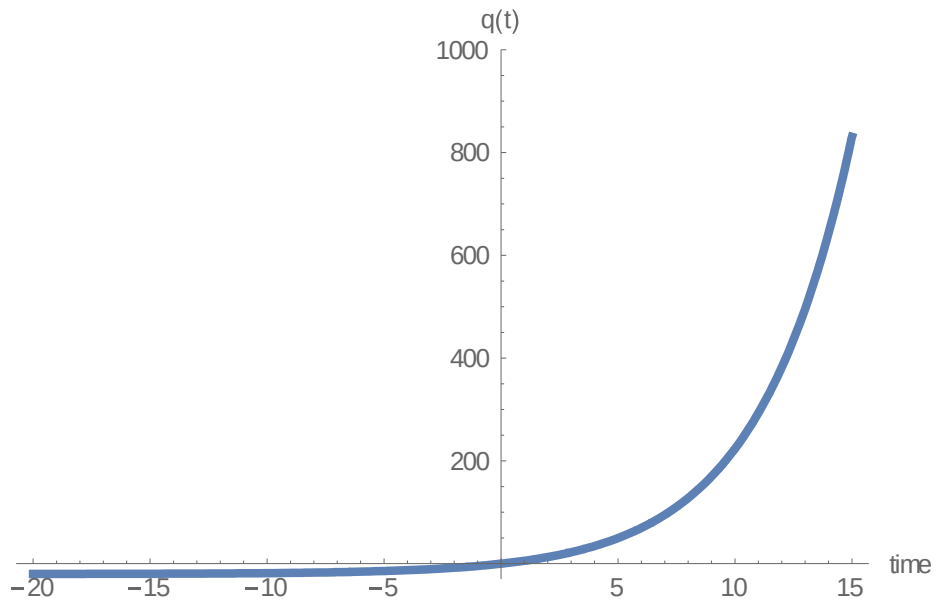

Figure 2a. Eq. (9) with $m = 2,\ a = 100,\ A = 90,\ B = -0.5,\ H_0 = 20.$

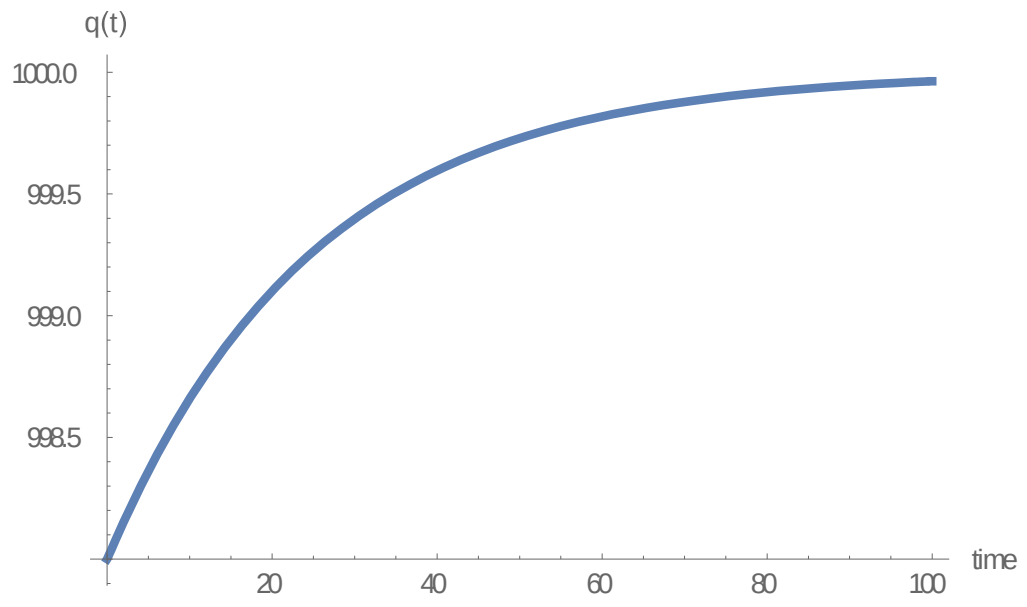

Figure 2b. Eq. (9) with $m = 2, a = 100,\ A = 20,\ B = 0.08,\ H_0 = -2.$

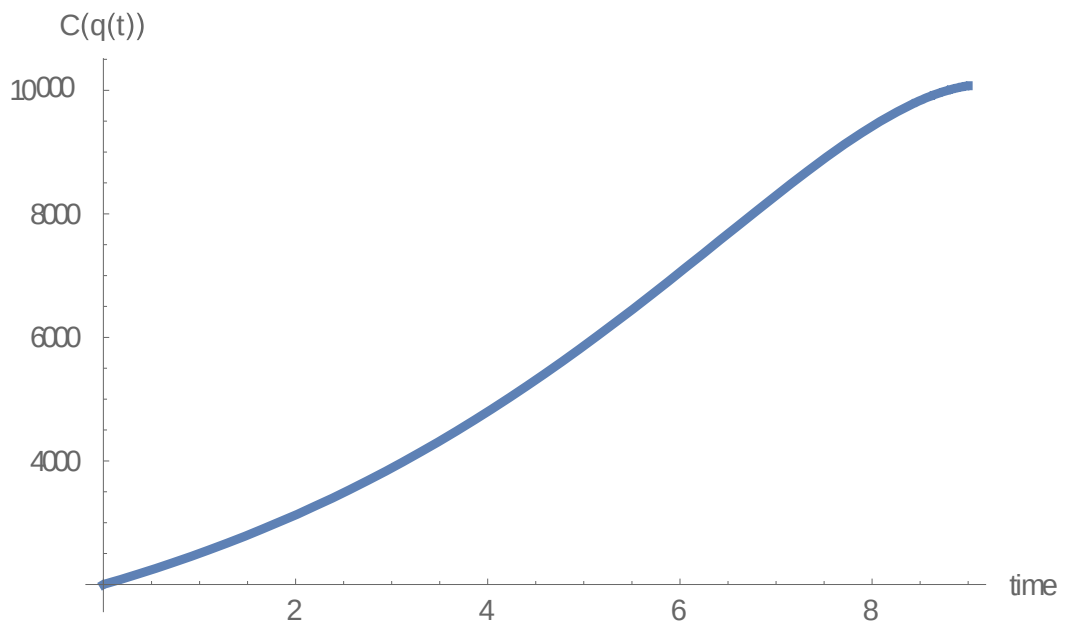

Figure 3a. Cost function in Eq. (5) with $m = 2$, $a = 100,\ A = 90,\ B = -0.5,\ h_0 = 2000,\ H_0 = 20.$

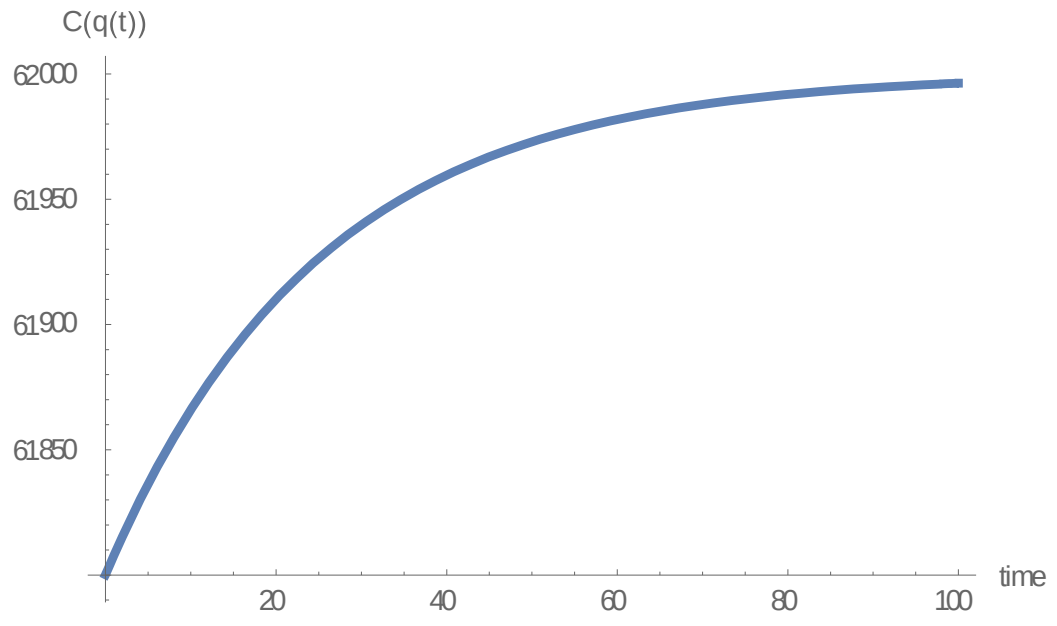

Figure 3b. Cost function in Eq. (5) with $m = 2$, $a = 100,\ A = 20,\ B = 0.08,\ h_0 = 2000,\ H_0 = -2.$

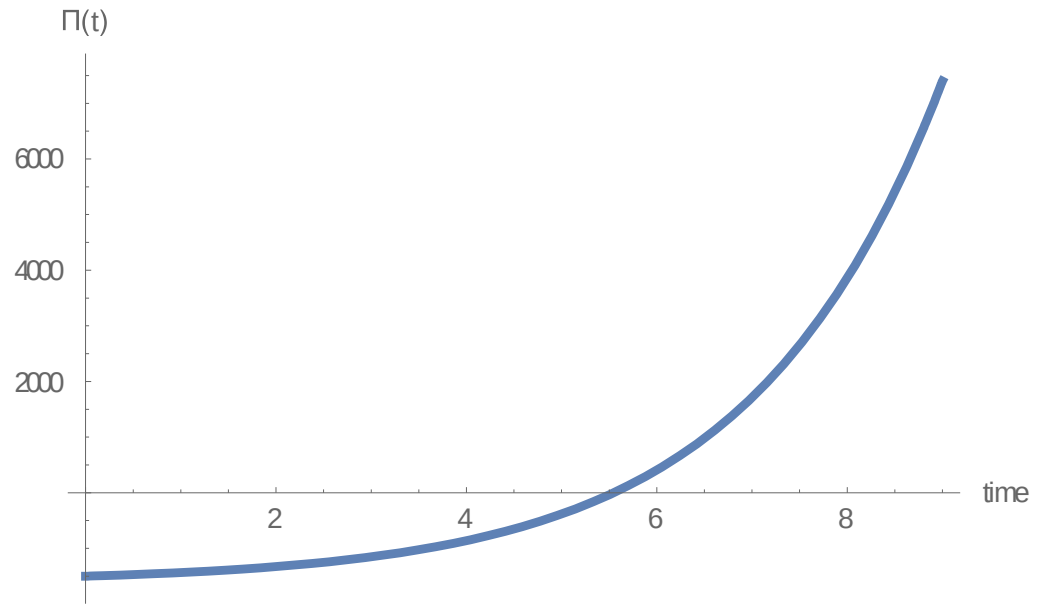

Figure 4a. Profit function in Eq. (6) with $m = 2,\ A = 90$, $a = 100,\ B = -0.5,\ h_0 = 2000,\ H_0 = 20.$

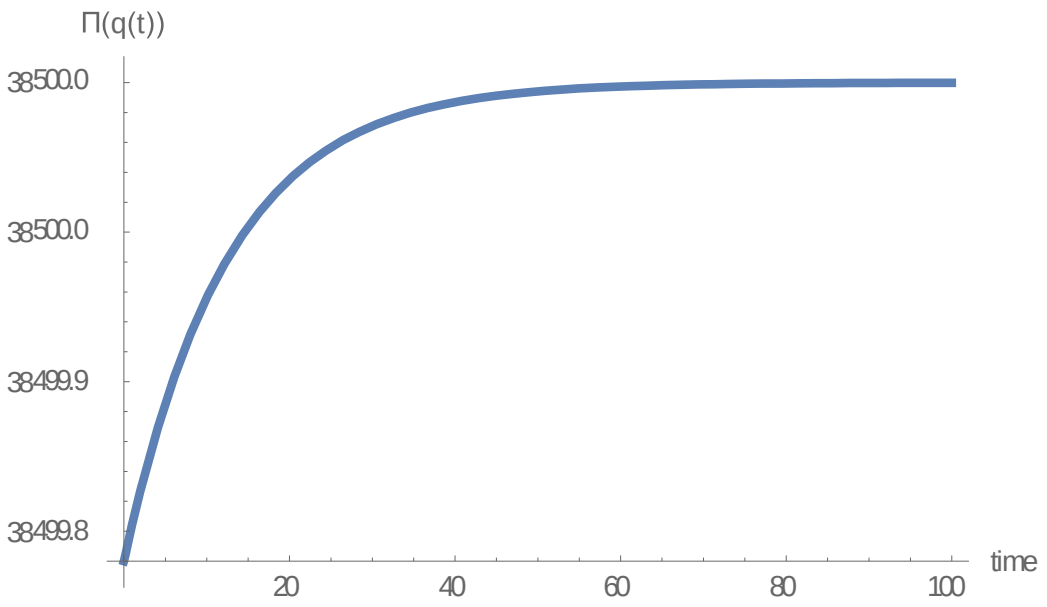

Figure 4b. Profit function in Eq. (6) with $m = 2,\ A = 20$, $a = 100,\ B = 0.08,\ h_0 = 2000,\ H_0 = -2.$

## 5.2. Physical interpretation of the model

The "free body" diagram of the system in Eq. (8) is in Fig. 5. There "mass" $m$ represents the "particle" called the accumulated production of the firm[3], on the horizontal axis is accumulated production, the direction of motion toward accumulating production is defined positive, $a - A$ is the constant positive driving force component, and $Bq$ the resisting force component that depends on the flow of production $q$. In the neoclassical optimum, "particle" $m$ moves with constant velocity $q^*$ to the direction where accumulated production increases. Because $q$ cannot be negative, the particle either moves to the "right" in the coordinate system or stays constant if $q = 0$. If $B < 0$, the resisting force component is positive, however, and then both force components affect in the positive direction of motion. This explains why in that case the flow of production has an exponential time trend.

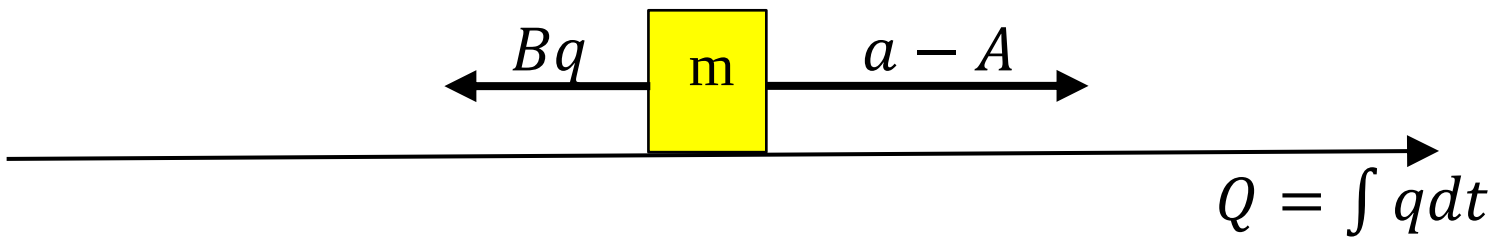

Figure 5. The forces acting upon the production of the firm

[3]The box-shape of the "particle" has no economic meaning, and actually the particle should be presented as a point on line $Q$. However, the box-shape reveals the similarity to particle dynamics in physics, see Fig. 6.

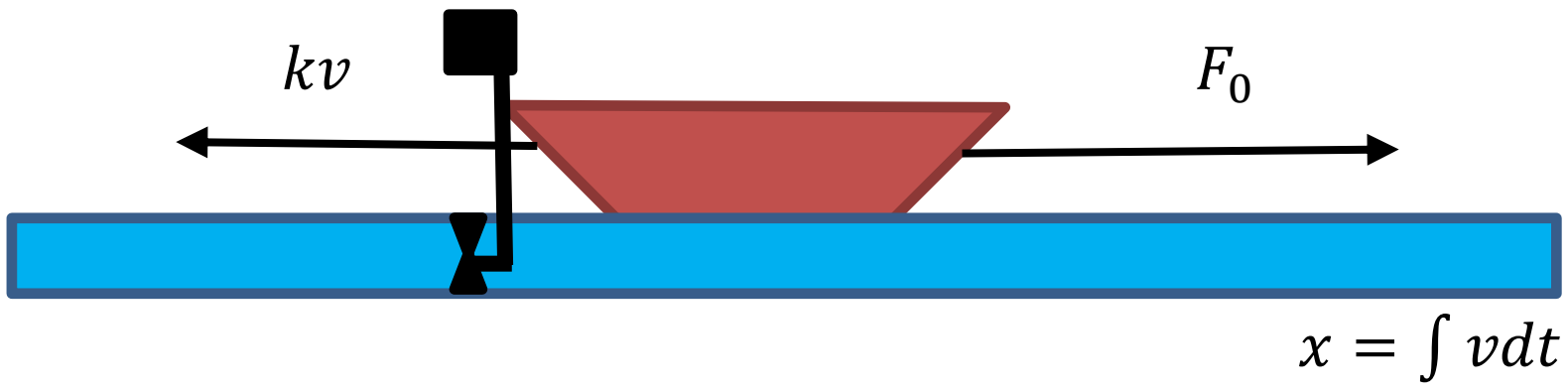

Figure 6. The forces acting upon a motorboat moving in water

The corresponding model in physics is e.g. that of a motorboat moving in water that causes kinetic friction to resist the motion, see Fig. 6. The engine causes a positive force component $F_0(N)$, $N = Newton = kg \times m \times s^{-2}$, on the boat, which is assumed constant for simplicity. The engine pushes the boat at a constant force and the kinetic friction of water causes a negative force component against the motion. In physics, the kinetic friction of the resisting medium can be approximated by term $kv$, where $v = x'(t)(m/s)$ is the velocity of the boat, $x$ its position coordinate, and $k(kg/s)$ a positive constant that depends on the viscosity of the medium (water). The resisting force component is small if $v$ is small and it increases linearly with the speed. If starting at rest, the acceleration of the boat is first positive but with a constant pushing force, the speed stabilizes to a constant after reaching condition $F_0 = kv$. In Fig. 6, we have omitted the vertical force components caused by gravitation that add up to zero because of the buoyancy force component caused by water on the boat.

The Newtonian equation of motion for the boat is

$$F_0 - kx'(t) = m_b x''(t) \quad \Leftrightarrow \quad F_0 - kv(t) = m_b v'(t), \tag{10}$$

where $x'(t) = v(t)$ and $m_b$ is the mass of the boat. The solution of the latter form of Eq. (10) gives the velocity of the boat as a function of time as follows

$$v(t) = \frac{F_0}{k} + C_0 e^{-\frac{k}{m_b}t}. \tag{11}$$

In Eq. (11) the equilibrium velocity is $v^* = F_0/k$, and depending on the initial condition, the positivity or negativity of $C_0$, $C_0 = v(0) - F_0/k$, the speed has either positive or negative acceleration toward the equilibrium velocity. This model is identical with that in Eq. (8) if

$$q(t) = v(t), \; m = m_b, \; a - A = F_0, \; B = k.$$

Thus, the mathematical form of the dynamic model about the production of a profit-seeking firm is identical with that of a particle moving in a resisting medium. Notice that the increasing returns to scale case in the boat example corresponds to $k < 0$, and then the boat has two pushing force components. This would correspond e.g. to a situation where the boat is moving down a river that flows faster than the boat. Then water pushes the boat forward and does not cause kinetic friction. The motion of the boat decreases to zero, exponentially, beginning at time moment $t_1$, if the pushing force is $\begin{cases} F_0, t \le t_1 \\ 0, t > t_1 \end{cases}$ with $k > 0$ (notice that setting $F_0 < 0$ after $t_1$ would speed up the stopping process). This kind of force could happen e.g. due to running out of gasoline at $t_1$. This corresponds to exponential decrease in the flow of production, i.e. the bankruptcy of the firm, if the driving force of production vanishes or gets negative, i.e. $a \le A$ with $B > 0$.

### 5.3. Dynamics of production due to technological change or changing wealth

We showed in the previous section that increasing returns to scale does not create permanent growth in the flow of production, because the benefits of decreasing unit costs with increasing $q(t)$ vanish before obtaining zero unit costs. Thus, we need other reasons to explain the permanent growth of a firm or its bankruptcy. Let the firm in the previous section have the sales and cost functions as

$$p(t) = a + \frac{b}{q(t)} + ct, \qquad C(q(t),t) = h(t) + \left(A + \frac{B}{2}q(t) - Gt\right)q(t), \tag{12}$$

where constant $c$ with unit $€ \times y/unit$ represents the change in popularity of this product with time among consumers; $c > 0$ implies increasing popularity and vice versa. Constant $G$ with unit $€ \times y/unit$ represents technological development so that $G > 0$ makes unit costs decreasing with time and vice versa. More complicated time trends are possible, of course, but they will complicate the resulting Newtonian equation. Suppose $c, G \neq 0$ and $B > 0$. The Newtonian equation is then

$$\frac{\partial \Pi}{\partial q} = mq'(t) \quad \Leftrightarrow \quad a - A - Bq(t) + (c + G)t = mq'(t) \tag{13}$$

with the following solution

$$q(t) = \frac{(a - A)B - (c + G)m}{B^2} + \left(\frac{c + G}{B}\right)t + H_0 e^{-\frac{B}{m}t}, \tag{14}$$

where $H_0$ is the constant of integration. Eq. (14) shows that independent of the sign of $H_0$, the exponential time trend vanishes with time ($B > 0$). However, the linear time trend $(c + G)t/B$ keeps $q(t)$ changing so that $c + G > 0$ causes a positive time trend, and vice versa. Thus, either increasing popularity of the good among consumers or a technological progress keep $q(t)$ increasing with time. Here we have to make the same remark as with increasing returns to scale; technological development cannot make unit costs negative, i.e. in this case too unlimited growth is not possible. However, no limit exists for consumers' behavior, and if consumers' wealth or the popularity of this product increase continuously ($c > 0$), permanent growth in the flow of production will result.

On the other hand, $c + G < 0$ will cause the firm into bankruptcy with time, and the time moment of bankruptcy can be solved from Eq. (16) setting $q(t) = 0$. No analytic solution exists for the time moment of bankruptcy, however, but we can show by numerical methods that increases in constants $a, m, c, G, B$ increase the survival time of the firm when it is going into bankruptcy ($c + G < 0, B > 0$), while increasing $A$ diminishes the survival time. We can test these results by observing the time trends of the sales of firms and their unit costs, if necessary data exists.

## 6. Conclusions

We introduced a dynamic framework for modeling a firm's production that can explain besides equilibrium also growth and bankruptcy. We defined the "economic force" acting upon the production of a profit-seeking firm, and the model yields the neo-classical theory as a special case: zero-force situation in a stable case. The mathematical form of the model for the production of a firm turned out to be identical with Newton's model of a particle moving in a resisting medium. However, in physics a plenty of tools to model dynamic phenomena has been invented that may be applicable in other sciences too. This creates a great potential for the development of economics.